\documentclass[twocolumn,floatfix,aps,superscriptaddress]{revtex4} 

\usepackage{amsmath} 
\usepackage{graphicx}
\usepackage{dcolumn}
\usepackage{bm}

\usepackage{graphicx}

\newcommand{\nuc}[2]{$^{#1}${#2}} 
\newcommand{\frc}[2]{\mbox{$\frac{#1}{#2}$}} 
 
\newcommand{\ie}{\mathrm{i}} 
\newcommand{\de}{\mathrm{d}} 
\begin{document}

\title{%
The role of the continuum and the spurious $1^-$ transitions in 
incoherent 
$\mu^- - e^- $ conversion 
rate calculations 
}



\author{P. Papakonstantinou} 
\email[Email:]{panagiota.papakonstantinou@physik.tu-darmstadt.de} 
\affiliation{Institut f\"ur Kernphysik, Technische Universit\"at Darmstadt, D-64289 Darmstadt, Germany} 

\author{O. Civitarese} 
\affiliation{Departamento de F\'isica, Universidad Nacional de La Plata, c.c. 67 1900 La Plata, Argentina}

\author{T.S. Kosmas} 
\affiliation{Department of Physics, University of Ioannina, GR-45110 Ioannina, Greece} 

\author{J. Wambach} 
\affiliation{Institut f\"ur Kernphysik, Technische Universit\"at Darmstadt, D-64289 Darmstadt, Germany}



\begin{abstract}
By using the Continuum RPA (CRPA) method, the incoherent 
transition strength of the exotic $\mu^- - e^- $ conversion in the 
$^{208}$Pb and $^{40}${Ca} nuclei is investigated. The question whether 
excited nuclear states lying high in the continuum give an important 
contribution to the incoherent rate is addressed. 
The admixture of spurious components in the rate coming from 
$1^-$ excitations is investigated in detail by using the 
self-consistent CRPA with Skyrme interactions as well as a less 
consistent version and by employing two ways to remove the spurious strength: 
the use of effective operators or simply the exclusion 
of the spurious state appearing close to zero energy. 
In all cases, the correction achieved is quite large. 
\end{abstract}


\maketitle

\section{Introduction}
\label{intro}
Theories that go beyond the Standard Model (SM) of particle physics 
allow for the violation of various symmetries respected by the SM, 
among which the conservation of lepton flavour. 
Evidence for lepton flavour violation (LVF) has been provided 
by the neutrino oscillation experiments. In the charged-lepton sector, 
processes that would provide additional evidence for LVF 
and help distinguish among the various proposed mechanisms, 
include the exotic neutrinoless conversion of 
a muon to an electron - where the neutrino and antineutrino involved in the conversion are 
assumed to be Majorana particles and can annihilate each other. 

In this context, the exotic conversion of a bound muon to an electron 
\begin{equation} 
\mu^- + (A,Z) \to e^- + (A,Z)^*  
\end{equation} 
has been studied both experimentally and theoretically~\cite{Schaaf,Molzon,Kuno,tsk01,KFV97,Okada,KV96,Suh-tsk,KFSV,SFK97}. 
The experimentally measured quantity 
is the ratio of the coherent rate, where the nucleus remains in its ground state, 
over the total capture rate. Although the coherent rate dominates the capture rate 
(accounting for about 90\% of it),  
in order to calculate this ratio and make meaningful comparisons with experiment, 
both the coherent and the incoherent rate need to be evaluated theoretically. 

Various methods have been employed for the calculation of the incoherent 
$\mu^- -  e^- $ conversion 
rate~\cite{tsk01,KV96,Suh-tsk,KFSV,SFK97,Chiang,Vehn,Pyatov,Civi-02}.  
State-by-state calculations performed 
within the shell model and the quasiparticle RPA (QRPA) indicate that 
\begin{enumerate} 
\item 
the main contribution to the incoherent rate comes from low-lying states 
\item 
the contribution of the $1^-$ channel is very large (about 50\%, for 
all mechanisms leading to   
$\mu^- -  e^- $ conversion). 
Therefore, it is essential to properly remove possible spurious 
center-of-mass (CM) contaminations.  
\end{enumerate} 

In a recent paper~\cite{PKWF05} 
we used a Continuum-RPA (CRPA) method with Skyrme interactions 
to address the question 
as to how insignificant the contribution 
of excited states lying high in the continuum really is. 
We considered natural-parity excitations of the \nuc{208}{Pb} nucleus 
and found that high-lying strength is not negligible. 
In this work we continue this investigation by looking at a lighter nucleus 
as well, namely \nuc{40}{Ca}.  
We also examine in detail the admixture of spurious components in the 
rate coming from $1^-$ excitations. To this end we have used both the 
self-consistent CRPA and a non-consistent version (ncRPA) and have 
employed two ways to remove the spurious strength: by using effective 
dipole operators, as in Ref.~\cite{PKWF05}, and  by simply throwing away the spurious state appearing close 
to zero energy. In both cases, the correction achieved is quite large.

Useful definitions and basic information on the methods are provided in Sec.~\ref{method}. 
In Sec.~\ref{resl} we apply our CRPA method, which takes the full continuum into account, 
to the nuclear targets $^{208}$Pb and \nuc{40}{Ca}. 
The spurious strength is discussed in detail in Sec.~\ref{spour}. 
We conclude in Sec.~\ref{concl}. 

\section{Definitions and method of calculation} 
\label{method} 

The inclusive $(\mu^-,e^-)$ rate is evaluated 
by summing the partial contribution of all final states $|f\rangle$. 
For spherical or nearly spherical nuclei, the vector contribution is given by~\cite{tsk01} 
\begin{equation} 
S_a=\sum_f\left(\frac{q_f}{m_{\mu}}\right)^2 |\langle f | 
O_a(\bm{q}_f) | 0 \rangle |^2 , 
\label{vec-con}  
\end{equation} 
where $O_a(\bm{q}_f)$ represents the 
vector-type transition operator
resulting in the context of a given mechanism mediated by a photon ($a=\gamma$),
a $W$-boson ($a=W$) or a $Z$-particle exchange ($a=Z$). 
Here $\bm{q}_f$, with magnitude  
$q_f=m_{\mu}-\epsilon_b-E_f$, 
is the momentum transferred to the nucleus. 
$E_f$ is the energy of the final state $|f\rangle$ with 
respect to the ground state $|0\rangle$, 
$\epsilon_b$ is the binding energy of the muon and $m_{\mu}$ its mass. 
The transition operators have the form 
\begin{equation} 
O_a(\bm{q}) =  \tilde{g}_Vf_V 
\sum_{j=1}^A 6c_a(\tau_{ j }) 
\mathrm{e}^{-\ie \bm{q}\cdot\bm{r}_j} ,  
\hspace{2mm} 
c_a(\tau_{ j }) \equiv \frc{1}{2}+\frc{1}{6}\beta_a\tau_{ j } 
,  
\label{Eiqrop}  
\end{equation} 
where $\tau_{ j }$ is the 3rd component of the $j$th particle's isospin. 
The parameter $f_V=1.0$ represents the vector static nucleon 
form factor and the normalization coefficient $\tilde{g}_V$ 
takes the value $1/6$ for the photonic case and $1/2$ for the 
non-photonic $W$ boson and SUSY $Z$ exchange \cite{KFV97}. 
The value of $\beta_a$ depends on the model assumed. 
We have adopted the values from Ref.~\cite{Suh-tsk}. 
Thus, protons (neutrons) contribute to 
a given process with a ``charge" whose 
value is determined by $c_{a}(1/2) = 1/2+\beta_a /6$ ($c_{a}(-1/2)=1/2-\beta_a /6$).  
In the photon and $Z$ case, the isoscalar and isovector components of 
the transition operator are (almost) equally important, 
whereas $O_W$ is predominantly isoscalar.

By assuming that the initial and final states are of definite spin
and parity, a multipole decomposition of the operators 
of Eq.~(\ref{Eiqrop}) 
into operators $T_{aLM}$ of orbital angular momentum rank $L$ 
can be carried out. For spherical nuclei we can assume, 
without loss of generality,  
$\hat{q}=\hat{z}$. Then, only terms with 
$M=0$ survive, for which we obtain 
($T_{aL}({q}) \equiv 
T_{aL0}({q})$)  
\begin{equation} 
T_{aL}({q}) 
=  \tilde{g}_Vf_V 
\sqrt{4\pi (2L+1)}  
\sum_{j=1}^A 6c_a(\tau_j) 
j_L(qr_j) Y_{L0} (\hat{r}_j) 
 . 
\label{Ejqrop}  
\end{equation} 
A phase factor $(-\textrm{i})^L$ has been omitted. 
The contribution of each multipolarity to 
the transition rate $S_a$ reads 
\begin{equation} 
S_{aL}=\sum_f \left( \frac{q_f}{m_{\mu}} \right)^2 
| \langle f | 
T_{aL}(q_f)|0\rangle |^2 . 
\end{equation} 
We now rewrite the rate $S_{aL}$ as the integral of 
a suitable distribution over excitation energy:  
\begin{equation} 
S_{aL}  \equiv \int\de E R_{aL}(E) 
\label{e:sint} 
\end{equation} 
with 
\begin{eqnarray} 
R_{aL}(E) &=&  
\left[ 
\frac{E^2}{m_{\mu}^2} 
-2k\frac{E}{m_{\mu}} 
+k^2  
\right] 
R'_{aL}(E).  
\end{eqnarray} 
In the above expression 
we have set  
$k\equiv 1 - \epsilon_b/m_{\mu}$,
while  
\begin{equation} 
R'_{aL}(E)= \sum_f 
| \langle f | 
T_{aL}( 
m_{\mu}-\epsilon_b-E_f   
)|0\rangle |^2 
\delta (E-E_f) 
\label{Espr}  
\end{equation}  
stands for the ``strength distribution" 
corresponding to the operator $T_{aL} (q)$, with  
$q = m_{\mu}-\epsilon_b-E$. 

The final states $|f\rangle$, excited by the single-particle operator $T_{aL}$, 
are of particle-hole ($ph$) type. 
Then, the distribution $R'_{aL}(E)$, 
and from it $R_{aL}(E)$, 
can be calculated 
following the standard RPA method. 
Subsequently, Eq.~(\ref{e:sint}) can be used to  
evaluate the total rate $S_{aL}$. 

We consider $ph$ excitations, built on top of the mean-field 
ground state of a closed-shell nucleus 
and subjected to the $ph$ 
residual interaction. 
In particular, the quantities introduced above 
are calculated 
using a self-consistent  
Skyrme-Hartree-Fock (SHF) plus Continuum-RPA (CRPA) model. 
The HF equations describing the ground state 
are derived variationally from the Skyrme energy functional. 
RPA excitations are considered on top of the HF ground state. 
The CRPA  
is formulated 
in coordinate space, so that the full particle continuum 
is taken into account. 
The same Skyrme interaction is used to calculate the 
ground state properties and the residual $ph$ interaction. 
The model is described in detail in Ref.~\cite{PKWF05} 
and refs. therein. 
In this work we have employed 
the SkM*\cite{BQB82} parametrization of the Skyrme force. 
It describes satisfactorily  giant resonances of stable nuclei, 
and therefore it is suitable for the present study. 
In order to test the sensitivity of our results on the 
interaction used, we have also used MSk7 \cite{GTP2001}, 
which has a large effective mass, thereby shifting 
most excited states to lower energies compared to the 
more reliable SkM*. 

For the purposes of Sec.~{\ref{spour}} we have 
employed also the non-consistent CRPA version (ncRPA) 
of Ref.~\cite{ShB1975} and the corresponding numerical code \cite{BeXX}. 
The ground state is described by a Woods-Saxon potential 
of radius $(A-1)^{1/3}r_0$ ($r_0=1.25$~fm) and diffuseness $a_0=0.65$~fm, 
including central (strength $V_0=-53$~MeV), spin-orbit ($V_{\mathrm{so}}=15.5$~MeV~fm$^2$), 
symmetry ($V_T=20$~MeV) and Coulomb terms. 
The residual $ph$ interaction is a simplified Skyrme interaction 
without spin and velocity dependence ($t_0=-1100~$MeV~fm$^3$, $x_0=0.5$, $t_3=15000$~MeV~fm$^6$). 
Its strength is scaled by a factor $V_{\mathrm{scal}}$ so as to to bring the spurious state 
close to zero energy. 
 
Results 
are presented
and discussed in the next section. 
 
\section{Results} 
\label{resl} 

Next, we present results for  
the nuclei \nuc{208}{Pb} and \nuc{40}{Ca}. 
The muon binding energy in \nuc{208}{Pb} is
$\epsilon_b = 10.475$~MeV. 
The particle threshold energy $E_{\mathrm{thr}}$ 
is 8.09~MeV 
in the case of the SkM* force. 
The respective values for \nuc{40}{Ca} are $\epsilon_b=1.0533$~MeV, 
$E_{\mathrm{thr}}=8.86$~MeV. 
We have obtained results for $L=0,1,\ldots,6$ and 
for natural parity, $(-1)^L$. 
The most important contributions to the 
incoherent transition rate are expected from 
$L<4$ \cite{SFK97}.  

\subsection{Incoherent transition rate in the continuum} 

\begin{figure*} 
\includegraphics[height=0.22\textheight]{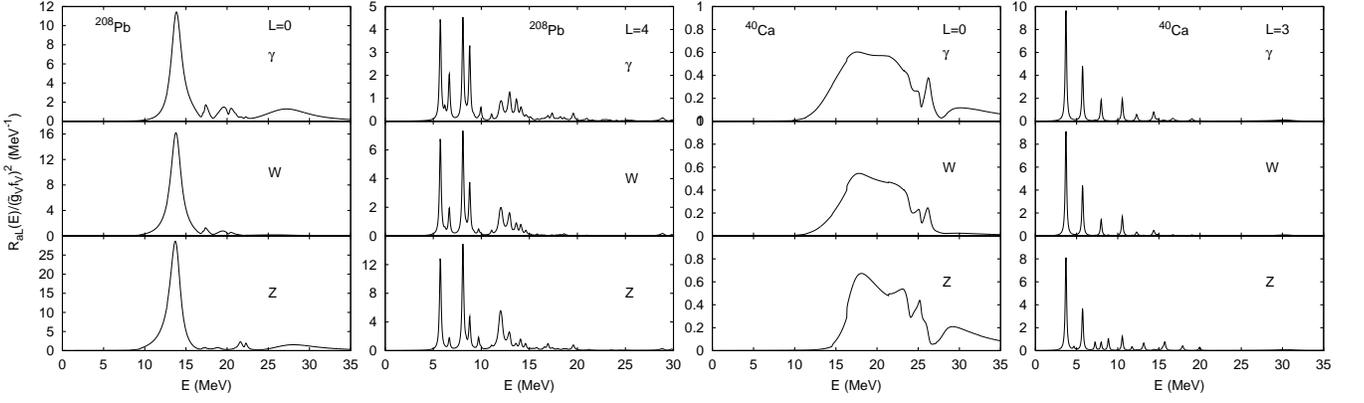} 
\caption{%
\label{Fstot1a}  
The distribution $R_{aL}(E)$ in $^{208}$Pb for $L=0,4$ and in \nuc{40}{Ca} for $L=0,3$. 
Skyrme parameterization SkM* has been used.}  
\end{figure*}

In Fig.~\ref{Fstot1a}, first two panels, 
the distribution $R_{aL}(E)$ is plotted as a function of $E$, 
for the $0^+$ and $4^+$ transitions 
of \nuc{208}{Pb}. 
For the $1^-$ distribution of \nuc{208}{Pb} the reader is referred to 
Fig.~\ref{Fspur} (first panel, full lines)
in the next subsection. 
In the monopole case, $L=0$, 
the Isoscalar (IS) Giant Monopole Resonance (GMR)  is the main peak. 
For $\gamma$ and Z, there is considerable contribution coming
from higher energies (20-35~MeV), i.e., the isovector (IV) 
GMR region. 
For $L=1$, 
the IV Giant Dipole Resonance (GDR) 
corresponds to the strength 
clustered around $E \approx 12$~MeV. 
For $W$ and $Z$, important contribution seems to come 
from higher energies 
(above 20~MeV), in particular, the IS GDR. 
For $W$ exchange, the region below 10 MeV contributes 
significantly. In this region we find the oscillation 
of the neutron skin against the nuclear core 
(pygmy dipole resonance)~\cite{SIS90}. 
In the quadrupole case, $L=2$ (not shown), 
both the IS Giant Quadrupole Resonance (GQR), 
close to 11~MeV, 
and the collective 
low-lying state are strong. 
There is some contribution from energies higher than 
15~MeV, i.e., from the IV GQR region, 
especially in the cases $\gamma$  and $Z$. 
For $L=3$ (not shown) the strength is mostly 
concentrated in the collective octupole state 
at low energy. 
For $L>3$, eg. for $L=4$, 
the calculated strength is quite fragmented and 
most of it lies below 20~MeV. 


In Fig.~\ref{Fstot1a}, last two panels, 
the distribution $R_{aL}(E)$ is plotted as a function of $E$, 
for the $0^+$ and $3^-$ transitions 
of \nuc{40}{Ca}. 
For the $1^-$ distribution of \nuc{40}{Ca} the reader is referred to 
Fig.~\ref{Fspur} (second panel, full lines)
in the next subsection. 
The monopole strength distribution is dominated by the 
broad IS GMR. Above 27~MeV the IV GMR is excited through the $\gamma$ and $Z$ mechanisms. 
The IV GDR dominates the dipole spectrum. 
In the $2^+$ distribution we find the strong IS GQR 
at about 17~MeV
and seemingly little strength at higher energies.  
The $3^-$, $4^+$ (not shown) distributions are quite fragmented. 
A strong peak appears at about 5~MeV in the $5^-$ distribution (not shown)  
along with frargmented strength at higher energies. 
In the $6^+$ distribution (not shown) we find a strong peak at around 15~MeV and 
a broad structure in the continuum between 20 and 30~MeV.

\begin{figure} 
\includegraphics[angle=-90,width=0.98\columnwidth]{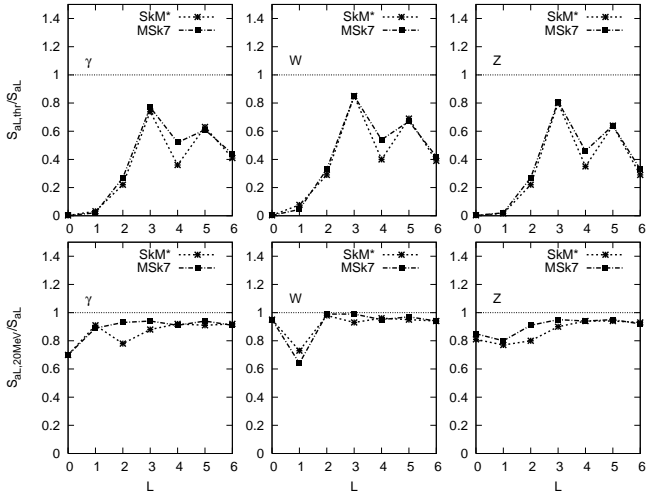} 
\caption{%
\label{Fs20THa}  
Fraction of the total strength $S_{aL}$, for the nucleus \nuc{208}{Pb}, 
coming from states below the 
particle threshold (top) and below 20~MeV (bottom) vs the multipolarity $L$. 
Skyrme parameterizations SkM* and MSk7 have been used.
Lines are drawn to guide the eye. 
}  
\end{figure}

In Fig.~\ref{Fs20THa} we plot the fraction of the total strength $S_{aL}$ 
coming from states below the particle threshold 
($S_{aL,\mathrm{thr}}$) 
and the fraction coming from states below 20~MeV 
($S_{aL,20\mathrm{MeV}}$),  
vs. the multipolarity $L$, 
for \nuc{208}{Pb} and using the Skyrme parametrizations 
SkM* and MSk7. 
We see that for low multipoles $L=0,1,2$ only a small portion of the
strength originates from energies below particle threshold.
The trend followed is similar for all mechanisms
and 
independent of the interaction used. For even multipoles $L=2,4$
a big portion of the contribution is pushed to higher energies as
compared to the neighboring odd ones. 
For some multipoles ($L=0$ for photonic mechanism,
$L=1$ for $W$-boson exchange), a significant portion of the strength
comes from above $20$MeV.
%


\begin{figure} 
\includegraphics[angle=-90,width=0.98\columnwidth]{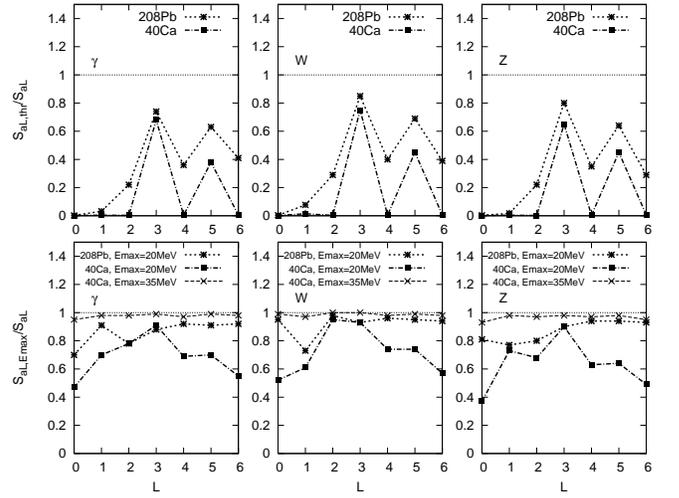} 
\caption{%
\label{Fs20THb}  
Fraction of the total strength $S_{aL}$, for the nuclei \nuc{208}{Pb} and \nuc{40}{Ca}, 
coming from states below the 
particle threshold (top) and below 20~MeV (bottom) vs the multipolarity $L$. 
For \nuc{40}{Ca} the fraction of strength coming from 
below 35~MeV is also shown (bottom).  
Skyrme parameterization SkM* has been used. 
Lines are drawn to guide the eye. 
}  
\end{figure}

In Fig.~\ref{Fs20THb} we show the fractions $S_{aL,\mathrm{thr}}/S_{aL}$ 
and $S_{aL,20\mathrm{MeV}}/S_{aL}$ 
for both nuclei \nuc{208}{Pb} and \nuc{40}{Ca}, evaluated using the 
SkM* force. We also show, for \nuc{40}{Ca}, the fraction $S_{aL,35MeV}/S_{aL}$ coming 
from states below 35~MeV.  
The quantities $S_{aL,\mathrm{thr}}/S_{aL}$ 
for \nuc{40}{Ca} show that 
for $L=0,1$ and, in addition, for $L=2,4,6$ 
almost all of the strength comes from above threshold. One should bare in mind 
that \nuc{40}{Ca} is an $\ell -$ closed nucleus, 
without low-lying $\Delta N=0$ $ph$ states. 
\nuc{208}{Pb}, on the other hand, is $\ell j-$ closed and low-lying 
$L=2,4,6$ transitions between spin-orbit partners are present.  
For odd multipolarities in \nuc{40}{Ca} an important amount of strength 
comes from continuum excitations as well. 
A larger fraction of strength is found above 20~MeV for \nuc{40}{Ca} 
than for \nuc{208}{Pb}. 
In this sense,  
high-lying excitations are found more important for 
\nuc{40}{Ca} than for \nuc{208}{Pb}. 
Of course, the value of $E_{\max}=20$~MeV 
was chosen arbitrarily. 
The oscillator level spacing 
($\hbar\omega \approx 41A^{-1/3}$) of a light nucleus 
is larger than for a heavier nucleus and the energies of the 
corresponding excitations are higher. 
If we choose to compare, e.g.,  
the $S_{aL,20MeV}/S_{aL}$ of \nuc{208}{Pb} with the $S_{aL,35MeV}/S_{aL}$ of 
\nuc{40}{Ca} (the maximum energy being $\approx3$~shells in both cases), we find that, in the latter 
case, values closer to unity are reached. 
 

We have calculated also the  fraction of the total strength $S_{aL}$, 
coming from states below 50~MeV, for $L$ up to 6 and for both nuclei. 
In all cases, the fraction is practically equal to unity. 
This means that the discretized versions of RPA and QRPA 
are safe to use if 
the energy cutoff is large enough to sufficiently account for transitions 
below this value.

\subsection{Removal of spurious strength} 
\label{spour}

It is well known \cite{Vehn,Pyatov} that the $1^-$ excitations contain 
admixtures of the spurious excitation of the center of mass (CM) of the nucleus,
corresponding to a situation in which the nucleus moves as a 
whole around the localized fictitious potential well. Normally, 
spurious components are separated out by the RPA methods. However, the 
use of a truncated model space and non-self-consistent single particle 
energies in ordinary RPA and the other versions of QRPA destroys 
the translational invariance and inserts spurious
excitations into the spectrum. Thus, the spurious CM 
state is not completely separated from the real (intrinsic) nuclear 
excitations, and in addition its energy eigenvalue is not zero. 
In Continuum-RPA models with Skyrme interactions it has been possible 
to achieve a high degree of self-consistency, i.e. the same interaction 
is used for the HF calculation of ground state properties and for the 
residual interaction. In addition, no truncation is involved. However, due 
to the formulation of the model in coordinate space, it is 
common practice to exclude the Coulomb and spin-orbit contribution 
(at least) to the residual interaction. Therefore, self-consistency is 
violated and, even in cases where the spurious state appears very close 
to zero energy, some spurious strength may remain at higher energies. 

For electric dipole excitations, the problem is usually treated by using effective 
charges \cite{BM69}. 
Similarly, in the case of IS dipole excitations, effective operators 
are used \cite{VaS1981a,ASS03}, 
which minimize the spurious admixture in the strength 
distribution. 
(The effect on the IS dipole excitations of \nuc{208}{Pb} was examined in detail in 
Ref.~\cite{HaS2002b}.)  
In Ref.~\cite{PKWF05} we presented a similar prescription 
for the operators involved 
in $\mu^- - e^-$ conversion. 
In particular, the operators $T_{a1}$ which induce the $1^-$ excitations,  
\begin{equation} 
\label{Eomg} 
T_{a1}(q)=\sum_{j=1}^{A}c(\tau_j)f(r_j) 
Y_{10}(\hat{r}_j)  
; f(r) = 6\tilde{g}_Vf_V\sqrt{12\pi}j_1(qr), 
\end{equation} 
are replaced by respective effective operators 
\begin{equation} 
T_{a1}^{\mathrm{corr}}(q)=\sum_{j=1}^{A}[c(\tau_j)f(r_j) -\eta_a r_j] 
Y_{10}(\hat{r}_j) 
\label{Eomgcorr} 
. 
\end{equation} 
In principle, the operators $T_{a1}$ and $T_{a1}^{\mathrm{corr}}$ 
induce the same intrinsic excitations, because they only differ by a term which translates the 
center of mass. 
The parameter $\eta_a$ in Eq.~(\ref{Eomgcorr}) is determined 
so as to eliminate the spurious CM excitation within the collective model. 
One finds~\cite{PKWF05} 
\begin{equation} 
\eta_a = \tilde{g}_Vf_V 4\sqrt{3\pi} q 
\left[ \frac{c_{ap}Z}{A} F_p(q) + \frac{c_{an}N}{A} F_n(q)  \right] 
. 
\label{Eetaa}  
\end{equation} 
The point-proton and neutron form factors, 
$F_p(Q)$ and $F_n(Q)$ respectively, 
are calculated numerically using the ground-state densities. 

\begin{figure*} 
\includegraphics[height=0.22\textheight]{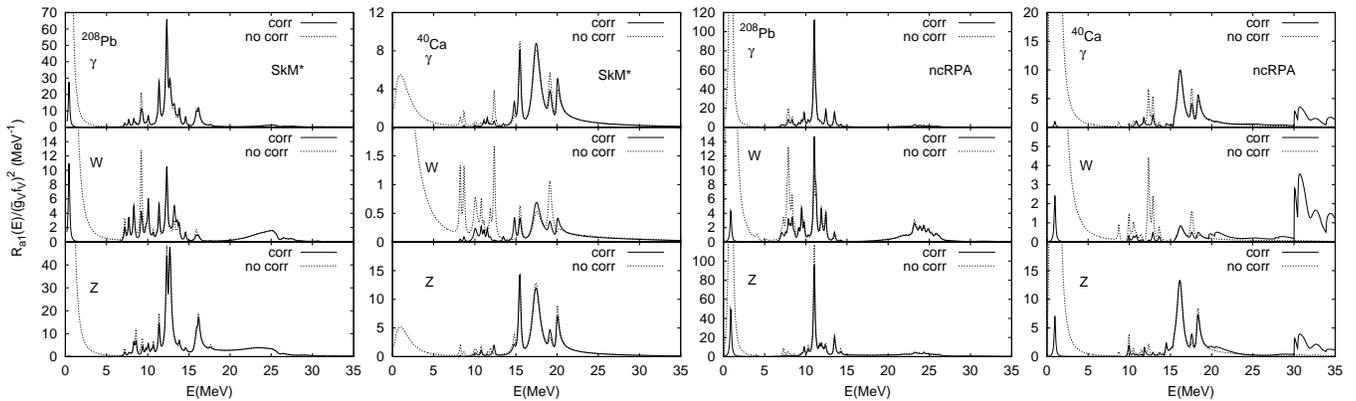} 
\caption{%
\label{Fspur}  
The dipole distributions $R_{a1}(E)$, for $\gamma$-photon and $W$-boson
exchange diagrams of the $\mu^-\to e^-$ conversion in $^{208}$Pb and $^{40}$Ca. The
results have been calculated for the dipole operator $T_{a1}$ (dotted line) 
and for 
the corresponding 
corrected operator given by Eqs. (\ref{Eomgcorr}), (\ref{Eetaa}). 
First two panels: The self-consistent CRPA with Skyrme parametrization SkM* was used. 
Last two panels: the non-self-consistent RPA was used. 
}  
\end{figure*}

In Fig.~\ref{Fspur}, first two panels, we plot the dipole distributions $R_{a1}(E)$ of 
Eq.~(\ref{Espr}), for photon, $W$- and $Z$-boson exchange diagrams,
calculated by using the corrected and uncorrected operator. 
The SkM* interaction was used. Results are shown for both \nuc{208}{Pb} and \nuc{40}{Ca}. 
Most of the spurious strength 
below $\approx$6~MeV has been removed. 
The strength distributions above 20~MeV are practically unaffected. 
The strength in the region of the IVD resonance 
appeares redistributed. The effect of the correction 
appears strongest in the case of the 
$W$-boson exchange mechanism and for \nuc{40}{Ca}. 
The pygmy dipole state of \nuc{208}{Pb} below 10~MeV 
is strongly affected by the correction 
in the photonic and $W$ cases.


In most of the cases, the effect of the correction above the tail of the spurious 
state appears small. Therefore, it would have been a fairly good approximation 
to simply throw away the spurious state and calculate the transition rate coming from 
states above, approximately, 6~MeV. 
Such an approximation was used in Ref.~\cite{SFK97}. One should keep in mind, however, that 
our CRPA model is self-consistent to a high degree. Thus, the excitation spectrum 
above the spurious state is almost free of 
spurious components, already before the operator correction, and therefore almost insensitive (up to 10~\%, 
as we will see) to 
the change of operators. This is not necessarily the case in less consistent models 
such as the one in Ref.~\cite{SFK97} and the ncRPA model described in Sec.~\ref{method}. 

Next we employ the ncRPA method and calculate again the $R_{a1}(E)$ distributions 
using the operators $T_{a1}$ and $T_{a1}^{\mathrm{corr}}$. 
In the case of \nuc{208}{Pb} (\nuc{40}{Ca}) the residual interaction was 
scaled by a factor  $V_{\mathrm{scal}}= 0.94~(1.004)$ in order to bring the 
spurious state close to zero energy. The results are shown in Fig.~\ref{Fspur}, 
last two panels.  
Again, in all cases, most of the strength of the spurious state is eliminated. 
In general, the effect appears small for \nuc{208}{Pb}, above the tail of the spurious state. 
The pygmy dipole state is strongly affected and the strength of the IS GDR is slightly 
reduced (see $W-$boson mechanism). 
For \nuc{40}{Ca} the effect appears small in the region of the IV GDR, but not 
below and above. In particular, the use of corrected operators results in 
significant additional strength above 20~MeV. In total, the transition 
strength above the spurious state (above 6~MeV) increases. The sensitivity of the 
calculation on the operators used implies a bad degree of consistency which 
renders the calculation unreliable.  
 

In Table~\ref{table1} we list, for both nuclei, 
i) the portion $s^{\mathrm{sp}}_{\mathrm{tot}}$ of transition strength removed from  
the total contaminated $1^-$ transition strength $S_{a 1}$ when corrected operators 
$T_{\mathrm{a1}}^{\mathrm{corr}}$ are 
employed, 
ii) the portion of strength $S_{\mathrm{sp}}$ carried 
by the spurious state when the uncorrected $T_{a1}$ operators are used 
(calculated as the portion of strength lying below 6~MeV) 
and iii) the portion 
$s^{\mathrm{sp}}_{>6{\mathrm{MeV}}}$ 
of the strength removed from above 6~MeV excitation energy 
when corrected operators are used  
(with respect to the uncorrected 
strength above 6~MeV).  
Results have been obtained with both models, i.e., our 
self-consistent CRPA model with interaction SkM* and the non-self consistent 
CRPA model labelled ncRPA. 

Let us first examine the behavior of the self-consistent CRPA 
model. As the values of $s_{\mathrm{tot}}^{\mathrm{sp}}$ show, 
about 90\% of the total transition rate 
was spurious in all cases. 
We expect this result to be independent of the nucleus and the interaction used, 
because the spurious state at low energy  
always dominates the isoscalar dipole strength distribution 
(which contributes in all three mechanisms) and because the 
corrected operators are, by construction,  
most effective for this state (thus removing practically 
all its strength). 
We were not able to demonstrate this in the particular case 
of \nuc{40}{Ca}, 
because the energy of the 
spurious state in this case was found imaginary. 
In other words, we were not able to evaluate 
and take into account properly the strength of the spurious state, 
before or after the correction. 
Therefore, the respective numbers in Table~\ref{table1} are placed in brackets. 

Suppose now that, 
in order to evaluate the rate coming 
from intrinsic $1^-$ excitations, 
we would apply the procedure of using the $T_{a1}$ operators and 
then simply excluding the strength of the spurious 
state from calculating the total rate. 
As the values of $S_{\mathrm{sp}}$ show, for \nuc{208}{Pb}, we would have removed roughly as large a 
portion as $s_{\mathrm{tot}}^{\mathrm{sp}}$. 
Again, for \nuc{40}{Ca} it is not possible to conclude.  

From Table~\ref{table1} we notice that for $s^{\mathrm{sp}}_{>6{\mathrm{MeV}}}$ 
and for $a=\gamma$ the result is small in absolute value, but negative, 
for \nuc{208}{Pb}. 
It represents the numerical accuracy of our calculation and, being small, 
it indicates that the degree of self consistency reached by our HF+CRPA model 
is sufficient to achieve a satisfactory separation of the spurious 
transition in this case. 
For the $W$ and $Z$ cases, however, spurious admixtures of more than 6\% 
are found above 6~MeV. 
These numbers 
(which are not free of numerical inaccuracies, as explained in Ref.~\cite{PKWF05})  
vary when different Skyrme interactions are used, with their 
values remaining below 10\%. 
In the case of \nuc{40}{Ca}, 
the values of $s_{\mathrm{>6MeV}}^{\mathrm{sp}}$ for the $\gamma$ and $Z$ 
mechanisms are lower than 10\% as well. Its value for $W$-exchange is much larger. 
Most of it, however, comes from below 15~MeV (below the GDR region).

When the ncRPA model is used, similar results are obtained for 
$s_{\mathrm{tot}}^{\mathrm{sp}}$ 
and $S_{\mathrm{sp}}$ 
as with the self-consistent model. 
The \nuc{208}{Pb} excitation spectrum above the tail of the spurious state 
is a little more contaminated than within the self-consistent CRPA, as the  
$s_{\mathrm{>6MeV}}^{\mathrm{sp}}$ 
values show. 
As for the \nuc{40}{Ca} nucleus, when corrected operators were used, the strength 
above 6~MeV was significantly increased. Thus, the 
corresponding values of  
$s_{\mathrm{>6MeV}}^{\mathrm{sp}}$ 
are large and negative, indicating a bad performance of the model. 

From our results, obtained for two nuclei within two different continuum-RPA 
models, one may conclude that 
more than 85\% of the total $1^-$ rate, when no correction is considered, 
is spurious. 
Whether effective operators are used to remove the total spurious strength, 
or whether one excludes the spurious state from the calculation of 
the total $1^-$, the above conclusion remains the same. 
This does not mean that there are no spurious contaminations above the 
zero-energy spurious mode. 
When making use of corrected operators, the strength of the excited states across the 
spectrum is redistributed. 
The overall effect, however, is to reduce the total $S_{a1}$ rate by the 
amount that was initially carried by the spurious state. 

Within the QRPA calculations 
of  Ref.~\cite{SFK97}, for six nuclei (including \nuc{208}{Pb}), 
a more moderate correction (less than 60\%) on $S_{a1}$ was achieved 
by considering the 
lowest $1^-$ state as purely spurious and simply excluding it. 
One should keep in mind the differences between the models 
used there (QRPA calculations with a renormalized G-matrix interaction) 
and the ones used here. First of all, priority was given 
in Ref.~\cite{SFK97} to 
the best possible reproduction of experimental spectra, rather than 
controlling the self-consistency. 
In addition, a truncated model space was used, contrary to the methods used 
in the present work. 
The various approximations entering 
may have influenced the spuriosity results in an unpredictable way.  
Both consistency and completeness of 
the space 
are important in order to move as much spurious strength as possible 
close to zero energy. Of course, the lowest 
$1^-$ state appeared very close to zero energy. It was found that it was 
spurious by 60-80\% (except for \nuc{126}{Yb}, for which it was 
about 90\% spurious). Thus, the approximation of considering it as purely spurious 
is not a very good one. The numbers mentioned above refer to the overlap of the 
lowest $1^-$ state with the ``purely spurious" RPA state. 
Approximations were inevitable when constructing the 
purely spurious state, in order to normalize it.



\begin{table} 
\begin{center} 
\begin{tabular}{lr|ccc}   
\hline 
\hline 
 & &  $\gamma$ & $W$  & $Z$      \\ 
\hline 
$s^{\mathrm{ sp}}_{\mathrm{ tot}}(\% )$    & \nuc{208}{Pb} - SkM* &  86.9    & 96.3 & 90.5   \\  
                                          & \nuc{208}{Pb} - ncRPA &  89.8    & 96.5 & 88.5   \\ 
                                            & \nuc{40}{Ca} - SkM* & [33.8]   & [82.0] & [28.6] \\  
                                           & \nuc{40}{Ca} - ncRPA & 90.8 & 93.7 & 88.0  \\  
\hline  
$S_{\mathrm{ sp}}  (\% )                $ & \nuc{208}{Pb} - SkM* &  88.0    & 96.6 & 89.9    \\  
                                           & \nuc{208}{Pb} - ncRPA &  90.0    & 96.5 & 90.3    \\  
                                            & \nuc{40}{Ca} - SkM* & [29.0]   & [67.7] & [21.6] \\  
                                           & \nuc{40}{Ca} - ncRPA & 95.2 & 98.7 & 93.6  \\  
\hline  
$s^{\mathrm{ sp}}_{>6{\mathrm{ MeV}}}(\% )$ & \nuc{208}{Pb} - SkM* &  -1.3    & 7.8  & 6.1     \\  
                                           & \nuc{208}{Pb} - ncRPA &  -1.6    & 12.5 & 16.2    \\  
                                            & \nuc{40}{Ca} - SkM* &   6.7    & 45.0 & 9.1     \\  
                                           & \nuc{40}{Ca} - ncRPA &  incr.   & incr. & incr.   \\  
\hline 
\hline 
\end{tabular} 
\end{center} 
\caption{%
\label{table1} 
Percentage 
of the total $1^-$ transition strength $S_{a 1}$ ($s^{\mathrm{ sp}}_{\mathrm{ tot}}$) 
and of the strength above 6~MeV ($s^{\mathrm{sp}}_{>6{\mathrm{ MeV}}}$)  
consumed by 
spurious transitions, 
and percentage of strength below 6~MeV ($\approx$~strength of the spurious state, $S_{\mathrm{sp}}$) 
for 
the three channels $\gamma$, $W$, $Z$. 
Results are shown for the nuclei \nuc{208}{Pb} and \nuc{40}{Ca} 
obtained by using two models: the self-consistent Continuum RPA with Skyrme force 
SkM* and the non-consistent RPA (ncRPA). 
Last row: When corrected operators are used in this case, the strength above 6~MeV increases 
significantly.  
Square brackets: the energy of the spurious state is found imaginary, 
therefore its strength is not properly evaluated.  
}  
\end{table}

\section{Conclusions} 
\label{concl} 

We have investigated the incoherent rate of the
exotic $\mu^- - e^-$ conversion in the nuclei 
$^{208}$Pb 
and 
\nuc{40}{Ca}. We employed
the Continuum-RPA method which is appropriate for
explicit construction of the excited states lying in the continuum spectrum of the
nuclear target. 
We used a self-consistent CRPA and Skyrme interactions  
to investigate the transition strength 
coming from natural-parity $ph$ excitations up to $L=4$. 
We found that a significant portion of the
incoherent $\mu^- - e^-$ rate comes from high-lying nuclear excitations. 
The admixture of spurious components in the rate coming from 
$1^-$ excitations was investigated in detail by using the 
self-consistent CRPA with Skyrme interactions as well as a less 
consistent version of CRPA and by employing two ways to remove the spurious strength: 
the use of effective operators or simply the exclusion 
of the spurious state appearing close to zero energy. 
In all cases, we found that the greatest portion of the $1^-$ transition strength is due
to the spurious CM excitation, a result in agreement with that of an exact method 
constructed recently \cite{Civi-02} for removing spurious 
contaminations.

\bigskip 
{\small 
This work has been supported by the 
Deutsche For\-schungs\-ge\-mein\-schaft 
under contract SFB 634 
and 
by the 
Hellenic General Secretariat for 
Research and Technology (www.gsrt.gr) under Program PENED-03 
(Space Sciences and Technologies). 
T.S.K. acknowledges partial support from the 
IKYDA-02 Greece-Germany project. 
P.P. acknowledges support from the 
ILIAS-N6 EU project for participation to the MEDEX-05 meeting. 
} 



\end{document}